\documentclass[review]{elsarticle}

\usepackage{lineno,hyperref,color}
\modulolinenumbers[5]

\journal{Materials Science in Semiconductor Processing}

\begin{document}

\begin{frontmatter}

\title{Bulk moduli of PbS$_{x}$Se$_{1-x}$, PbS$_{x}$Te$_{1-x}$, and PbSe$_{x}$Te$_{1-x}$ from the combination of the $cB\Omega$ model with the modified Born theory compared to  generalized gradient approximation.}

\author{Nicholas V. Sarlis}
\ead{nsarlis@phys.uoa.gr}

\author{Efthimios S. Skordas}
\ead{eskordas@phys.uoa.gr}

\address{Department of Solid State Physics and Solid Earth Physics
Institute, Faculty of Physics, School of Science, National and Kapodistrian University of Athens, Greece.}



\begin{abstract}
The bulk moduli of 
PbS$_{x}$Se$_{1-x}$, PbS$_{x}$Te$_{1-x}$, and PbSe$_{x}$Te$_{1-x}$  have been recently  determined 
[E. Skordas,  Materials Science in Semiconductor Processing 43 (2016) 65–68] by employing 
a thermodynamical model, the so called $cB\Omega$ model, which has been found to give successful results in several applications of defects in solids. Here, we suggest an alternative procedure for this determination which combines the $cB\Omega$ model with the modified Born theory. The results are in satisfactory agreement with those deduced independently by the generalized gradient approximation approach. 
\end{abstract}


\end{frontmatter}


\section{Introduction}

Since small fundamental energy band gap\cite{Dal74,Cow65} exists in lead chalcogenides semiconductors of rocksalt structure, it makes them very useful in optoelectronic equipment, including lasers and detectors\cite{Tac95,Fei96,Spr98,Kho03}, and very promising for the photoinduced nonlinear optics\cite{Kit04,Nou06}. Thus, they have attracted a major interest and have been studied by various theoretical techniques\cite{Lac02,Alb00,Alb05}. In this frame,  the structural, electronic and optical properties of PbS$_{x}$Se$_{1-x}$, PbS$_{x}$Te$_{1-x}$, and PbSe$_{x}$Te$_{1-x}$  ternary alloys have been studied recently for $x=0.25, 0.5, 0.75$ and $1.00$ by \citet{Nae14} by using the first-principle technique of full-potential linearized augmented plane-wave (FP-LAPW) method with Wu-Cohen generalized gradient approximation (GGA) \cite{Wu06} to solve Kohn-Sham equation\cite{Koh65}. In particular, concerning the structural properties \citet{Nae14} calculated the equilibrium lattice constant ($a$), the bulk modulus ($B$), and the pressure derivative of the bulk modulus ($\frac{dB}{dP}$) at the aforementioned compositions. In our previous publication\cite{Sko16}, focusing on the variation of $B$ versus the composition, the above GGA results were compared with those obtained on the basis of a thermodynamical model, termed  $cB\Omega$ model\cite{VARBOOK} (see also below),  which has been applied for the formation and migration of defects  to a variety of solids including 
several recent applications in semiconductors. Here, we extend our previous study and discuss an alternative procedure to determine $B$ versus the composition in the aforementioned semiconductors' alloys, which combines the thermodynamical $cB\Omega$ model with the modified Born theory of solids \cite{Roberts1970619}.

\section{The alternative procedure for the determination of the compressibility of PbS$_{x}$Se$_{1-x}$, PbS$_{x}$Te$_{1-x}$, and PbSe$_{x}$Te$_{1-x}$}
Let us explain how the compressibility $\kappa$ of a solid solution A$_{1-x}$B$_{x}$
is interrelated with compressibilities of the two end members A and B by following Refs. \cite{VARBOOK,SKO12}.
We call the two end members A and B as pure components 1 and 2, respectively, and label $v_1$ the volume per atom of the pure component 1 and  $v_2$ the volume per atom of the pure component 2. 
Let $V_1$  and $V_2$ denote the corresponding molar volumes, i.e. $V_1=Nv_1$  and $V_2=Nv_2$ (where $N$ stands for Avogadro's number) and assume that $v_1 < v_2$.
We now define a ``defect  volume'' \cite{VARBOOK,SKO12} as the increase of the volume $V_1$ if one atom of type 1 is replaced by one atom of type 2. 
It is now evident that the addition of one ``atom'' of type 2 to a crystal containing  atoms of type 1 will increase its volume by $v_d+v_1$ (see pp.325 and 326 of Ref. \cite{VARBOOK}). 
Assuming that $v^d$ is independent of composition, the volume $V_{N+n}$ of a crystal containing $N$  atoms of type 1 and  $n$ atoms of type 2 can be written as 
\begin{equation}\label{eq2}
V_{N+n}=Nv_1+n(v^d+v_1) \Longleftrightarrow V_{N+n}= \left[ 1+\frac{n}{N} \right] V_1+nv^d .
\end{equation}
The molar fraction $x$ is connected to $n/N$ by (see Eq.(12.5) on p.328 of Ref.\cite{VARBOOK})
\begin{equation}\label{eq3}
\frac{n}{N}=\frac{x}{1-x} .
\end{equation}
The compressibility $\kappa$ of the solid solution (as well as its bulk modulus $B=1/\kappa$) can be found by differentiating Eq.(\ref{eq2}) with respect to
pressure, which finally gives: 
\begin{equation}\label{eq4}
\kappa V_{N+n}=\kappa_1V_1+\frac{n}{N} \left[ \kappa^d N v^d +\kappa_1 V_1 \right],
\end{equation}
where $\kappa^d$ denotes the compressibility of the volume $v^d$, defined as
\begin{equation}\label{eq5}
\kappa^d\equiv \frac{1}{B^d}=-\frac{1}{v^d} \left( \frac{\partial v^d}{\partial P} \right)_T.
\end{equation}
The ``defect volume'' $v^d$ can be approximated by (see p.342 of Ref.\cite{VARBOOK})
\begin{equation}\label{eq6}
v^d=\frac{V_2-V_1}{N} \Longleftrightarrow v^d=v_2-v_1.
\end{equation}

The quantity $V_{N+n}$ can be obtained versus the composition by means of Eq.(\ref{eq2})
when considering also Eq.(\ref{eq6}) and then, the compressibility $\kappa$ can be studied versus the composition from Eq.(\ref{eq4}) by making the crucial assumption that $\kappa^d$ is independent of composition. 
An estimation of $\kappa^d$ can be made by employing the aforementioned thermodynamical model\cite{VAR77A,VAR78B,VAR07} (for a review see Ref.\cite{VARBOOK}) for the formation and migration of the defects in solids (cf. the replacement of a host atom with a ``foreign'' one can be considered in general as a defect \cite{VAR74}). 
This model has been successfully applied, as mentioned, to various categories of solids including diamond\cite{PhysRevB.75.172107}, fluorides\cite{VALEX80,VAR819,VARALEX80,Varotsos2008438},
semiconductors\cite{CHR15A}, mixed alkali halides\cite{PSSB80,VAR81,VAR80AA,VALEX80B} as well as in complex ionic materials under uniaxial 
stress that emit electric signals before fracture\cite{VAR99,VAR92}, which is reminiscent of the electric signals observed before major earthquakes\cite{VAR84A,VAR84B,VAR91,VAR93,NAT09V,NAT08,NEWTSA,CHAOS2010}.
This thermodynamical model states that the defect Gibbs energy $g$ is proportional to the bulk modulus as well as to the mean volume per 
atom. We then obtain the defect volume $v[=(\partial g /\partial P)_T]$, and therefrom the compressibility $\kappa^d$ (see Eq.(8.31) on p.156  of Ref.\cite{VARBOOK}) is found to be
\begin{equation}\label{eq7}
\kappa^d=\frac{1}{B} - \frac{ \left( \frac{\partial^2 B}{\partial P^2} \right)_T } {\left( \frac{\partial B}{\partial P} \right)_T-1},
\end{equation} 
where the quantity $\left(\frac{\partial^2 B}{\partial P^2}\right)_T$ can be approximated from the modified Born theory\cite{Roberts1970619} according to 
\begin{equation}\label{born}
\left(\frac{\partial^2 B}{\partial P^2}\right)_T=- \frac{4}{9} \frac{(n^B+3)}{B}
\end{equation}
where $n^B$ is the usual Born exponent. Upon inserting Eq.(\ref{born}) into Eq.(\ref{eq7}) we find $\kappa^d$ which can be replaced in Eq.(\ref{eq4}) leading to 
\begin{equation} \label{12.13}
\kappa=\kappa_1 \left\{ \frac{V_1}{V_{N+n}} \frac{(N+n)}{N}+\frac{n}{N} \frac{Nv^d}{V_{N+n}} \left[1+\frac{4}{9} \frac{(n^B+3)}{(\frac{dB_1}{dP}-1)} \right] \right\}.
\end{equation}
By taking into account that within Born approximation $\frac{dB_1}{dP}$ is given by\cite{VARBOOK,Roberts1970619}  
\begin{equation}
\frac{dB_1}{dP}=\frac{n^B+7}{3},
\end{equation}
we obtain
\begin{equation}
\kappa=\kappa_1 \left\{ \frac{V_1}{V_{N+n}} \frac{(N+n)}{N}+\frac{n}{N} \frac{Nv^d}{V_{N+n}} \left[1+\frac{4}{3} \frac{(n^B+3)}{(n^B+4)} \right] \right\}.
\end{equation}
Using now Eq.(\ref{eq6}), we have
\begin{equation}
\kappa=\kappa_1 \left\{ \frac{V_1}{V_{N+n}} \frac{(N+n)}{N}+\frac{n}{N} \frac{(V_2-V_1)}{V_{N+n}} \left[1+\frac{4}{3} \frac{(n^B+3)}{(n^B+4)} \right] \right\},
\end{equation}
which upon substituting $v_{N+n}=v_{N+n}/(N+n)$ leads to 
\begin{equation}
\kappa=\kappa_1 \left\{ \frac{v_1}{v_{N+n}} +\frac{n}{n+N} \frac{(v_2-v_1)}{v_{N+n}} \left[1+\frac{4}{3} \frac{(n^B+3)}{(n^B+4)} \right] \right\}
\end{equation}
or
\begin{equation}\label{eqres}
B=B_1 \left\{ \frac{v_1}{v_{N+n}} +x \frac{(v_2-v_1)}{v_{N+n}} \left[1+\frac{4}{3} \frac{(n^B+3)}{(n^B+4)} \right] \right\}^{-1}
\end{equation}
by virtue of Eq.(\ref{eq3}) and the definitions of the compressibilities $\kappa$ and $\kappa_1$.

Using Eq.(\ref{eqres}) and the values of $B_1$, $v_1$,  $v_2$ and $n^B$, e.g., $n^B=10.5$ for PbS and $n^B=11.0$ for PbSe (see p.30 of Ref.\cite{Sei40}; see also Fig.5-4 of \citet{anderson89}),   we can estimate $B$ as a function of the composition $x$. Adopting the stoichiometric notation PbS$_{1-x}$Se$_{x}$, PbS$_{1-x}$Te$_{x}$, and PbSe$_{1-x}$Te$_{x}$ we can compare the results of Eq.(\ref{eqres}) with those obtained by GGA \cite{Nae14}. Such a comparison is shown in Fig.1, an inspection of which reveals more or less a satisfactory agreement especially if one considers the following difference between the procedure presented here from the one suggested by \citet{Sko16}: Within the frame of the procedure followed in our previous paper, which is based solely on the thermodynamical model, in order to determine the bulk modulus of the alloy at a certain composition, one needs the bulk moduli of both end members, while in the frame of the present procedure, which in addition makes use of the modified Born theory,  one needs only the bulk modulus of the end member that has the smaller mean volume per atom.

\begin{figure}
\includegraphics[scale=0.72]{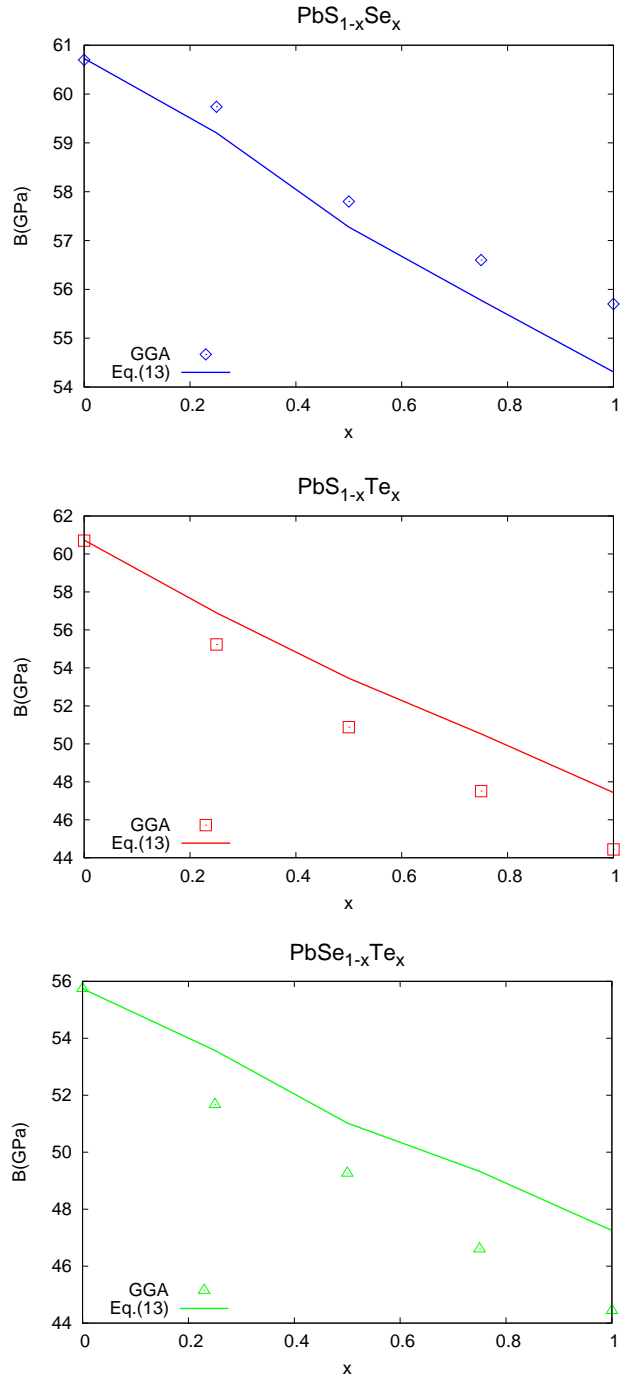}
\caption{Bulk modulus versus composition. The continous curve results from the present procedure, i.e., Eq.(\ref{eqres}), while the open symbols have been obtained by GGA.} 
\end{figure}

\section{Conclusions}
By combining the thermodynamical $cB\Omega$ model with the modified Born theory of solids, a procedure is proposed which enables the computation of the bulk moduli of the semiconductors' alloys PbS$_{1-x}$Se$_{x}$, PbS$_{1-x}$Te$_{x}$, and PbSe$_{1-x}$Te$_{x}$ at various compositions. Interestingly, the results are compatible with those that have been recently deduced by tedious GGA calculations.

\section*{References}


\begin{thebibliography}{44}
\expandafter\ifx\csname natexlab\endcsname\relax\def\natexlab#1{#1}\fi
\providecommand{\url}[1]{\texttt{#1}}
\providecommand{\href}[2]{#2}
\providecommand{\path}[1]{#1}
\providecommand{\DOIprefix}{doi:}
\providecommand{\ArXivprefix}{arXiv:}
\providecommand{\URLprefix}{URL: }
\providecommand{\Pubmedprefix}{pmid:}
\providecommand{\doi}[1]{\href{http://dx.doi.org/#1}{\path{#1}}}
\providecommand{\Pubmed}[1]{\href{pmid:#1}{\path{#1}}}
\providecommand{\bibinfo}[2]{#2}
\ifx\xfnm\relax \def\xfnm[#1]{\unskip,\space#1}\fi
\bibitem[{Dalven(1974)}]{Dal74}
\bibinfo{author}{R.~Dalven},
\newblock \bibinfo{title}{Electronic structure of pbs, pbse, and pbte},
\newblock volume~\bibinfo{volume}{28} of \textit{\bibinfo{series}{Solid State
  Physics}}, \bibinfo{publisher}{Academic Press}, \bibinfo{year}{1974}, pp.
  \bibinfo{pages}{179 -- 224}. \DOIprefix\doi{10.1016/S0081-1947(08)60203-9}.
\bibitem[{Cowley(1965)}]{Cow65}
\bibinfo{author}{R.~A. Cowley},
\newblock \bibinfo{title}{On the theory of ferroelectricity and anharmonic
  effects in crystals},
\newblock \bibinfo{journal}{Philos. Mag.} \bibinfo{volume}{11}
  (\bibinfo{year}{1965}) \bibinfo{pages}{673--706}.
  \DOIprefix\doi{10.1080/14786436508230077}.
\bibitem[{Tacke(1995)}]{Tac95}
\bibinfo{author}{M.~Tacke},
\newblock \bibinfo{title}{{New developments and applications of tunable IR lead
  salt lasers}},
\newblock \bibinfo{journal}{Infrared Phys. Technol.} \bibinfo{volume}{36}
  (\bibinfo{year}{1995}) \bibinfo{pages}{447 -- 463}.
  \DOIprefix\doi{10.1016/1350-4495(94)00101-P}. \bibinfo{note}{Proceedings of
  the Sixth International Conference on Infrared Physics}.
\bibitem[{Feit et~al.(1996)Feit, McDonald, Woods, Archambault, and Mak}]{Fei96}
\bibinfo{author}{Z.~Feit}, \bibinfo{author}{M.~McDonald},
  \bibinfo{author}{R.~J. Woods}, \bibinfo{author}{V.~Archambault},
  \bibinfo{author}{P.~Mak},
\newblock \bibinfo{title}{{Low threshold PbEuSeTe/PbTe separate confinement
  buried heterostructure diode lasers}},
\newblock \bibinfo{journal}{Appl. Phys. Lett.} \bibinfo{volume}{68}
  (\bibinfo{year}{1996}) \bibinfo{pages}{738--740}.
  \DOIprefix\doi{10.1063/1.116726}.
\bibitem[{Springholz et~al.(1998)Springholz, Holy, Pinczolits, and
  Bauer}]{Spr98}
\bibinfo{author}{G.~Springholz}, \bibinfo{author}{V.~Holy},
  \bibinfo{author}{M.~Pinczolits}, \bibinfo{author}{G.~Bauer},
\newblock \bibinfo{title}{{Self-Organized Growth of Three- Dimensional
  Quantum-Dot Crystals with fcc-Like Stacking and a Tunable Lattice Constant}},
\newblock \bibinfo{journal}{Science} \bibinfo{volume}{282}
  (\bibinfo{year}{1998}) \bibinfo{pages}{734--737}.
  \DOIprefix\doi{10.1126/science.282.5389.734}.
\bibitem[{Khoklov(2003)}]{Kho03}
\bibinfo{editor}{E.~Khoklov} (Ed.), \bibinfo{title}{Lead chalcogenides :
  physics and applications}, \bibinfo{publisher}{Taylor and Francis},
  \bibinfo{address}{New York}, \bibinfo{year}{2003}. \URLprefix
  \url{http://resolver.caltech.edu/CaltechBOOK:1989.001}.
\bibitem[{Kityk et~al.(2004)Kityk, Demianiuk, Majchrowski, Eboth\'{e}, and
  Siemion}]{Kit04}
\bibinfo{author}{I.~Kityk}, \bibinfo{author}{M.~Demianiuk},
  \bibinfo{author}{A.~Majchrowski}, \bibinfo{author}{J.~Eboth\'{e}},
  \bibinfo{author}{P.~Siemion},
\newblock \bibinfo{title}{{IR-induced second-harmonic generation in PbSe
  microcrystallites}},
\newblock \bibinfo{journal}{J. Phys.: Condens. Matter} \bibinfo{volume}{16}
  (\bibinfo{year}{2004}) \bibinfo{pages}{3533?3544}.
\bibitem[{Nouneh et~al.(2006)Nouneh, Kityk, Viennois, Benet, Charar, Paschen,
  and Ozga}]{Nou06}
\bibinfo{author}{K.~Nouneh}, \bibinfo{author}{I.~V. Kityk},
  \bibinfo{author}{R.~Viennois}, \bibinfo{author}{S.~Benet},
  \bibinfo{author}{S.~Charar}, \bibinfo{author}{S.~Paschen},
  \bibinfo{author}{K.~Ozga},
\newblock \bibinfo{title}{{Influence of an electron-phonon subsystem on
  specific heat and two-photon absorption of the semimagnetic semiconductors
  ${\mathrm{Pb}}_{1\ensuremath{-}x}{\mathrm{Yb}}_{x}X$
  $(X=\mathrm{S},\phantom{\rule{0.3em}{0ex}}\mathrm{Se},\mathrm{Te})$ near the
  semiconductor-isolator phase transformation}},
\newblock \bibinfo{journal}{Phys. Rev. B} \bibinfo{volume}{73}
  (\bibinfo{year}{2006}) \bibinfo{pages}{035329}.
  \DOIprefix\doi{10.1103/PhysRevB.73.035329}.
\bibitem[{Lach-hab et~al.(2002)Lach-hab, Papaconstantopoulos, and Mehl}]{Lac02}
\bibinfo{author}{M.~Lach-hab}, \bibinfo{author}{D.~A. Papaconstantopoulos},
  \bibinfo{author}{M.~J. Mehl},
\newblock \bibinfo{title}{{Electronic structure calculations of lead
  chalcogenides PbS, PbSe, PbTe}},
\newblock \bibinfo{journal}{J. Phys.Chem. Solids} \bibinfo{volume}{63}
  (\bibinfo{year}{2002}) \bibinfo{pages}{833 -- 841}.
  \DOIprefix\doi{10.1016/S0022-3697(01)00237-2}.
\bibitem[{Albanesi et~al.(2000)Albanesi, Okoye, Rodriguez, {Peltzer y Blanca},
  and Petukhov}]{Alb00}
\bibinfo{author}{E.~A. Albanesi}, \bibinfo{author}{C.~M.~I. Okoye},
  \bibinfo{author}{C.~O. Rodriguez}, \bibinfo{author}{E.~L. {Peltzer y
  Blanca}}, \bibinfo{author}{A.~G. Petukhov},
\newblock \bibinfo{title}{{Electronic structure, structural properties, and
  dielectric functions of IV-VI semiconductors: PbSe and PbTe}},
\newblock \bibinfo{journal}{Phys. Rev. B} \bibinfo{volume}{61}
  (\bibinfo{year}{2000}) \bibinfo{pages}{16589--16595}.
  \DOIprefix\doi{10.1103/PhysRevB.61.16589}.
\bibitem[{Albanesi et~al.(2005)Albanesi, y~Blanca, and Petukhov}]{Alb05}
\bibinfo{author}{E.~Albanesi}, \bibinfo{author}{E.~P. y~Blanca},
  \bibinfo{author}{A.~Petukhov},
\newblock \bibinfo{title}{{Calculated optical spectra of IV-VI semiconductors
  PbS, PbSe and PbTe}},
\newblock \bibinfo{journal}{Comput. Mat. Sci.} \bibinfo{volume}{32}
  (\bibinfo{year}{2005}) \bibinfo{pages}{85 -- 95}.
  \DOIprefix\doi{10.1016/j.commatsci.2004.07.001}.
\bibitem[{Naeemullah et~al.(2014)Naeemullah, Murtaza, Khenata, Hassan, Naeem,
  Khalid, and Omran}]{Nae14}
\bibinfo{author}{Naeemullah}, \bibinfo{author}{G.~Murtaza},
  \bibinfo{author}{R.~Khenata}, \bibinfo{author}{N.~Hassan},
  \bibinfo{author}{S.~Naeem}, \bibinfo{author}{M.~Khalid},
  \bibinfo{author}{S.~B. Omran},
\newblock \bibinfo{title}{{Structural and optoelectronic properties of
  PbS$_x$Se$_{1-x}$, PbS$_x$Te$_{1-x}$ and PbSe$_x$Te$_{1-x}$ via
  first-principles calculations}},
\newblock \bibinfo{journal}{Comput. Mat. Sci.} \bibinfo{volume}{83}
  (\bibinfo{year}{2014}) \bibinfo{pages}{496 -- 503}.
  \DOIprefix\doi{10.1016/j.commatsci.2013.10.033}.
\bibitem[{Wu and Cohen(2006)}]{Wu06}
\bibinfo{author}{Z.~Wu}, \bibinfo{author}{R.~E. Cohen},
\newblock \bibinfo{title}{More accurate generalized gradient approximation for
  solids},
\newblock \bibinfo{journal}{Phys. Rev. B} \bibinfo{volume}{73}
  (\bibinfo{year}{2006}) \bibinfo{pages}{235116}.
  \DOIprefix\doi{10.1103/PhysRevB.73.235116}.
\bibitem[{Kohn and Sham(1965)}]{Koh65}
\bibinfo{author}{W.~Kohn}, \bibinfo{author}{L.~J. Sham},
\newblock \bibinfo{title}{Self-consistent equations including exchange and
  correlation effects},
\newblock \bibinfo{journal}{Phys. Rev.} \bibinfo{volume}{140}
  (\bibinfo{year}{1965}) \bibinfo{pages}{A1133--A1138}.
  \DOIprefix\doi{10.1103/PhysRev.140.A1133}.
\bibitem[{Skordas(2016)}]{Sko16}
\bibinfo{author}{E.~Skordas},
\newblock \bibinfo{title}{{Bulk moduli of PbS$_x$Se$_{1-x}$, PbS$_x$Te$_{1-x}$
  and PbSe$_x$Te$_{1-x}$ from a thermodynamical model compared to generalized
  gradient approximation approach}},
\newblock \bibinfo{journal}{Mater. Sci. Semicond. Process.}
  \bibinfo{volume}{43} (\bibinfo{year}{2016}) \bibinfo{pages}{65 -- 68}.
  \DOIprefix\doi{10.1016/j.mssp.2015.12.002}; see also the relevant Corrigendum (in press).
\bibitem[{Varotsos and Alexopoulos(1986)}]{VARBOOK}
\bibinfo{author}{P.~Varotsos}, \bibinfo{author}{K.~Alexopoulos},
  \bibinfo{title}{Thermodynamics of Point Defects and their Relation with Bulk
  Properties}, \bibinfo{publisher}{North Holland},
  \bibinfo{address}{Amsterdam}, \bibinfo{year}{1986}.
\bibitem[{Roberts and Smith(1970)}]{Roberts1970619}
\bibinfo{author}{R.~Roberts}, \bibinfo{author}{C.~S. Smith},
\newblock \bibinfo{title}{Ultrasonic parameters in the born model of the sodium
  and potassium halides},
\newblock \bibinfo{journal}{J. Phys. Chem. Solids} \bibinfo{volume}{31}
  (\bibinfo{year}{1970}) \bibinfo{pages}{619 -- 634}.
  \DOIprefix\doi{10.1016/0022-3697(70)90196-4}.
\bibitem[{Skordas(2012)}]{SKO12}
\bibinfo{author}{E.~S. Skordas},
\newblock \bibinfo{title}{Comments on the elastic properties in solid solutions
  of silver halides},
\newblock \bibinfo{journal}{Mod. Phys. Lett. B} \bibinfo{volume}{26}
  (\bibinfo{year}{2012}) \bibinfo{pages}{1250066}.
  \DOIprefix\doi{10.1142/S0217984912500662}.
\bibitem[{Varotsos and Alexopoulos(1977)}]{VAR77A}
\bibinfo{author}{P.~Varotsos}, \bibinfo{author}{K.~Alexopoulos},
\newblock \bibinfo{title}{Calculation of the formation entropy of vacancies due
  to anharmonic effects},
\newblock \bibinfo{journal}{Phys. Rev. B} \bibinfo{volume}{15}
  (\bibinfo{year}{1977}) \bibinfo{pages}{4111--4114}.
  \DOIprefix\doi{10.1103/PhysRevB.15.4111}.
\bibitem[{Varotsos et~al.(1978)Varotsos, Ludwig, and Alexopoulos}]{VAR78B}
\bibinfo{author}{P.~Varotsos}, \bibinfo{author}{W.~Ludwig},
  \bibinfo{author}{K.~Alexopoulos},
\newblock \bibinfo{title}{{Calculation of the formation volume of vacancies in
  solids}},
\newblock \bibinfo{journal}{Phys. Rev. B} \bibinfo{volume}{18}
  (\bibinfo{year}{1978}) \bibinfo{pages}{2683--2691}.
  \DOIprefix\doi{10.1103/PhysRevB.18.2683}.
\bibitem[{Varotsos(2007)}]{VAR07}
\bibinfo{author}{P.~Varotsos},
\newblock \bibinfo{title}{Comparison of models that interconnect point defect
  parameters in solids with bulk properties},
\newblock \bibinfo{journal}{J. Appl. Phys.} \bibinfo{volume}{101}
  (\bibinfo{year}{2007}) \bibinfo{pages}{123503}.
\bibitem[{Varotsos and Mourikis(1974)}]{VAR74}
\bibinfo{author}{P.~Varotsos}, \bibinfo{author}{S.~Mourikis},
\newblock \bibinfo{title}{{Difference in conductivity between LiD and LiH
  crystals}},
\newblock \bibinfo{journal}{Phys. Rev. B} \bibinfo{volume}{10}
  (\bibinfo{year}{1974}) \bibinfo{pages}{5220--5224}.
\bibitem[{Varotsos(2007)}]{PhysRevB.75.172107}
\bibinfo{author}{P.~A. Varotsos},
\newblock \bibinfo{title}{Calculation of point defect parameters in diamond},
\newblock \bibinfo{journal}{Phys. Rev. B} \bibinfo{volume}{75}
  (\bibinfo{year}{2007}) \bibinfo{pages}{172107}.
  \DOIprefix\doi{10.1103/PhysRevB.75.172107}.
\bibitem[{Varotsos and Alexopoulos(1980)}]{VALEX80}
\bibinfo{author}{P.~Varotsos}, \bibinfo{author}{K.~Alexopoulos},
\newblock \bibinfo{title}{On the question of the calculation of migration
  volumes in ionic crystals},
\newblock \bibinfo{journal}{Philos. Mag. A} \bibinfo{volume}{42}
  (\bibinfo{year}{1980}) \bibinfo{pages}{13--18}.
  \DOIprefix\doi{10.1080/01418618008239352}.
\bibitem[{Varotsos and Alexopoulos(1981)}]{VAR819}
\bibinfo{author}{P.~Varotsos}, \bibinfo{author}{K.~Alexopoulos},
\newblock \bibinfo{title}{{Migration parameters for the bound fluorine motion
  in alkaline earth fluorides. II}},
\newblock \bibinfo{journal}{J. Phys. Chem. Sol.} \bibinfo{volume}{42}
  (\bibinfo{year}{1981}) \bibinfo{pages}{409 -- 410}.
  \DOIprefix\doi{10.1016/0022-3697(81)90049-4}.
\bibitem[{Varotsos and Alexopoulos(1980)}]{VARALEX80}
\bibinfo{author}{P.~Varotsos}, \bibinfo{author}{K.~Alexopoulos},
\newblock \bibinfo{title}{Migration entropy for the bound fluorine motion in
  alkaline earth fluorides},
\newblock \bibinfo{journal}{J. Phys. Chem. Sol.} \bibinfo{volume}{41}
  (\bibinfo{year}{1980}) \bibinfo{pages}{443--446}.
  \DOIprefix\doi{10.1016/0022-3697(80)90172-9}.
\bibitem[{Varotsos(2008)}]{Varotsos2008438}
\bibinfo{author}{P.~Varotsos},
\newblock \bibinfo{title}{{Point defect parameters in $\beta$-PbF$_{2}$
  revisited}},
\newblock \bibinfo{journal}{Solid State Ionics} \bibinfo{volume}{179}
  (\bibinfo{year}{2008}) \bibinfo{pages}{438 -- 441}.
  \DOIprefix\doi{10.1016/j.ssi.2008.02.055}.
\bibitem[{Chroneos and Vovk(2015)}]{CHR15A}
\bibinfo{author}{A.~Chroneos}, \bibinfo{author}{R.~Vovk},
\newblock \bibinfo{title}{Connecting bulk properties of germanium with the
  behavior of self- and dopant diffusion},
\newblock \bibinfo{journal}{Mater. Sci. Semicond. Process.}
  \bibinfo{volume}{36} (\bibinfo{year}{2015}) \bibinfo{pages}{179 -- 183}.
  \DOIprefix\doi{10.1016/j.mssp.2015.03.053}.
\bibitem[{Varotsos(1980)}]{PSSB80}
\bibinfo{author}{P.~Varotsos},
\newblock \bibinfo{title}{Determination of the dielectric constant of alkali
  halide mixed crystals},
\newblock \bibinfo{journal}{Phys. Status Solidi (b)} \bibinfo{volume}{100}
  (\bibinfo{year}{1980}) \bibinfo{pages}{K133--K138}.
\bibitem[{Varotsos(1981)}]{VAR81}
\bibinfo{author}{P.~Varotsos},
\newblock \bibinfo{title}{Determination of the composition of the maximum
  conductivity or diffusivity in mixed alkali halides},
\newblock \bibinfo{journal}{J. Phys. Chem. Sol.} \bibinfo{volume}{42}
  (\bibinfo{year}{1981}) \bibinfo{pages}{405--407}.
  \DOIprefix\doi{10.1016/0022-3697(81)90048-2}.
\bibitem[{Varotsos(1980)}]{VAR80AA}
\bibinfo{author}{P.~Varotsos},
\newblock \bibinfo{title}{On the temperature variation of the bulk modulus of
  mixed alkali halides},
\newblock \bibinfo{journal}{Phys. Status Solidi B} \bibinfo{volume}{99}
  (\bibinfo{year}{1980}) \bibinfo{pages}{K93--K96}.
  \DOIprefix\doi{10.1002/pssb.2220990243}.
\bibitem[{Varotsos and Alexopoulos(1980)}]{VALEX80B}
\bibinfo{author}{P.~Varotsos}, \bibinfo{author}{K.~Alexopoulos},
\newblock \bibinfo{title}{Prediction of the compressibility of mixed alkali
  halides},
\newblock \bibinfo{journal}{J. Phys. Chem. Sol.} \bibinfo{volume}{41}
  (\bibinfo{year}{1980}) \bibinfo{pages}{1291 -- 1294}.
  \DOIprefix\doi{10.1016/0022-3697(80)90130-4}.
\bibitem[{Varotsos et~al.(1999)Varotsos, Sarlis, and Lazaridou}]{VAR99}
\bibinfo{author}{P.~V. Varotsos}, \bibinfo{author}{N.~V. Sarlis},
  \bibinfo{author}{M.~S. Lazaridou},
\newblock \bibinfo{title}{Interconnection of defect parameters and
  stress-induced electric signals in ionic crystals},
\newblock \bibinfo{journal}{Phys. Rev. B} \bibinfo{volume}{59}
  (\bibinfo{year}{1999}) \bibinfo{pages}{24--27}.
  \DOIprefix\doi{10.1103/PhysRevB.59.24}.
\bibitem[{Varotsos et~al.(1992)Varotsos, Bogris, and Kyritsis}]{VAR92}
\bibinfo{author}{P.~Varotsos}, \bibinfo{author}{N.~Bogris},
  \bibinfo{author}{A.~Kyritsis},
\newblock \bibinfo{title}{Comments on the depolarization currents stimulated by
  variations of temperature and pressure},
\newblock \bibinfo{journal}{J. Phys. Chem. Solids} \bibinfo{volume}{53}
  (\bibinfo{year}{1992}) \bibinfo{pages}{1007--1011}.
  \DOIprefix\doi{10.1016/0022-3697(92)90069-P}.
\bibitem[{Varotsos and Alexopoulos(1984{\natexlab{a}})}]{VAR84A}
\bibinfo{author}{P.~Varotsos}, \bibinfo{author}{K.~Alexopoulos},
\newblock \bibinfo{title}{{Physical Properties of the variations of the
  electric field of the earth preceding earthquakes, I}},
\newblock \bibinfo{journal}{Tectonophysics} \bibinfo{volume}{110}
  (\bibinfo{year}{1984}{\natexlab{a}}) \bibinfo{pages}{73--98}.
  \DOIprefix\doi{10.1016/0040-1951(84)90059-3}.
\bibitem[{Varotsos and Alexopoulos(1984{\natexlab{b}})}]{VAR84B}
\bibinfo{author}{P.~Varotsos}, \bibinfo{author}{K.~Alexopoulos},
\newblock \bibinfo{title}{{Physical Properties of the variations of the
  electric field of the earth preceding earthquakes, II}},
\newblock \bibinfo{journal}{Tectonophysics} \bibinfo{volume}{110}
  (\bibinfo{year}{1984}{\natexlab{b}}) \bibinfo{pages}{99--125}.
  \DOIprefix\doi{10.1016/0040-1951(84)90060-X}.
\bibitem[{Varotsos and Lazaridou(1991)}]{VAR91}
\bibinfo{author}{P.~Varotsos}, \bibinfo{author}{M.~Lazaridou},
\newblock \bibinfo{title}{{Latest aspects of earthquake prediction in Greece
  based on Seismic Electric Signals}},
\newblock \bibinfo{journal}{Tectonophysics} \bibinfo{volume}{188}
  (\bibinfo{year}{1991}) \bibinfo{pages}{321--347}.
  \DOIprefix\doi{10.1016/0040-1951(91)90462-2}.
\bibitem[{Varotsos et~al.(1993)Varotsos, Alexopoulos, and Lazaridou}]{VAR93}
\bibinfo{author}{P.~Varotsos}, \bibinfo{author}{K.~Alexopoulos},
  \bibinfo{author}{M.~Lazaridou},
\newblock \bibinfo{title}{{Latest aspects of earthquake prediction in Greece
  based on Seismic Electric Signals,II}},
\newblock \bibinfo{journal}{Tectonophysics} \bibinfo{volume}{224}
  (\bibinfo{year}{1993}) \bibinfo{pages}{1--37}.
  \DOIprefix\doi{10.1016/0040-1951(93)90055-O}.
\bibitem[{Varotsos et~al.(2009)Varotsos, Sarlis, and Skordas}]{NAT09V}
\bibinfo{author}{P.~A. Varotsos}, \bibinfo{author}{N.~V. Sarlis},
  \bibinfo{author}{E.~S. Skordas},
\newblock \bibinfo{title}{Detrended fluctuation analysis of the magnetic and
  electric field variations that precede rupture},
\newblock \bibinfo{journal}{CHAOS} \bibinfo{volume}{19} (\bibinfo{year}{2009})
  \bibinfo{pages}{023114}. \DOIprefix\doi{10.1063/1.3130931}.
\bibitem[{Varotsos et~al.(2008)Varotsos, Sarlis, Skordas, and
  Lazaridou}]{NAT08}
\bibinfo{author}{P.~A. Varotsos}, \bibinfo{author}{N.~V. Sarlis},
  \bibinfo{author}{E.~S. Skordas}, \bibinfo{author}{M.~S. Lazaridou},
\newblock \bibinfo{title}{Fluctuations, under time reversal, of the natural
  time and the entropy distinguish similar looking electric signals of
  different dynamics},
\newblock \bibinfo{journal}{J. Appl. Phys.} \bibinfo{volume}{103}
  (\bibinfo{year}{2008}) \bibinfo{pages}{014906}.
  \DOIprefix\doi{10.1063/1.2827363}.
\bibitem[{Sarlis et~al.(2010)Sarlis, Skordas, and Varotsos}]{NEWTSA}
\bibinfo{author}{N.~V. Sarlis}, \bibinfo{author}{E.~S. Skordas},
  \bibinfo{author}{P.~A. Varotsos},
\newblock \bibinfo{title}{Nonextensivity and natural time: The case of
  seismicity},
\newblock \bibinfo{journal}{Phys. Rev. E} \bibinfo{volume}{82}
  (\bibinfo{year}{2010}) \bibinfo{pages}{021110}.
  \DOIprefix\doi{10.1103/PhysRevE.82.021110}.
\bibitem[{Skordas et~al.(2010)Skordas, Sarlis, and Varotsos}]{CHAOS2010}
\bibinfo{author}{E.~S. Skordas}, \bibinfo{author}{N.~V. Sarlis},
  \bibinfo{author}{P.~A. Varotsos},
\newblock \bibinfo{title}{Effect of significant data loss on identifying
  electric signals that precede rupture estimated by detrended fluctuation
  analysis in natural time},
\newblock \bibinfo{journal}{CHAOS} \bibinfo{volume}{20} (\bibinfo{year}{2010})
  \bibinfo{pages}{033111}. \DOIprefix\doi{10.1063/1.3479402}.
\bibitem[{Seitz(1940)}]{Sei40}
\bibinfo{author}{F.~Seitz}, \bibinfo{title}{The Modern Theory of Solids},
  \bibinfo{publisher}{McGraw-Hill Book Company}, \bibinfo{address}{New York},
  \bibinfo{year}{1940}.
\bibitem[{Anderson(1989)}]{anderson89}
\bibinfo{author}{D.~L. Anderson}, \bibinfo{title}{Theory of the Earth},
  \bibinfo{publisher}{Blackwell Scientific Publications},
  \bibinfo{address}{Boston}, \bibinfo{year}{1989}. \URLprefix
  \url{http://resolver.caltech.edu/CaltechBOOK:1989.001}.

\end{thebibliography}

\end{document}